# Altered Oscillatory Brain Networks During Emotional Face Processing in ADHD: An eLORETA–fICA Study

Saghar Vosough[5], Gian Candrian[1], Johannes Kasper[2], Hossam Abdel Rehim[3], Dominique Eich[4], Andreas Müller[1], Lutz Jäncke[5].

[1] Brain and trauma foundation Grisons/Switzerland, Chur, Switzerland

[2] Praxisgemeinschaft für Psychiatrie und Psychotherapie, Lucerne, Switzerland

[3] Psychiatrie und Psychotherapie Rapperswil, Rapperswil, Switzerland

[4] Department of Psychiatry, Psychotherapy, and Psychosomatics, University of Zurich, Zurich, Switzerland

[5] Division Neuropsychology, Department of Psychology, University of Zurich, Zurich, Switzerland

 **Corresponding author**: Saghar Vosough, Division of Neuropsychology, Department of Psychology, University of Zurich, Zurich, Switzerland.

Email: saghar.vosough@psychologie.uzh.ch

## Abstract

Attention-deficit/hyperactivity disorder (ADHD) is characterized by executive dysfunction and difficulties in processing emotional facial expressions, yet the large-scale neural dynamics underlying these impairments remain insufficiently understood.

This study applied network-based EEG source analysis to examine oscillatory cortical activity during cognitive and emotional Go/NoGo tasks in individuals with ADHD. EEG data from 272 participants (ADHD: n = 102; controls: n = 170; age 6–60 years) were analyzed using exact low-resolution brain electromagnetic tomography (eLORETA) combined with functional independent component analysis (fICA), yielding ten frequency-resolved cortical networks. Mixed-effects ANCOVAs were conducted on independent component (IC) loadings with Group, Task, and Condition as factors and age and sex as covariates. ADHD participants showed statistically significant but small increases in activation across several networks, including a gamma-dominant inferior temporal component (IC3) that exhibited both a Group effect and a Group × Condition interaction, reflecting disproportionately stronger NoGo-related activation in ADHD. Two additional components (IC6, IC9) showed similar but weaker NoGo-selective patterns. A Task main effect emerged only for IC8 (VCPT > ECPT). No Group × Task interactions were observed. Behavioral results replicated the established ADHD performance profile, with slower responses, greater variability, and higher error rates—particularly during the emotional ECPT. Overall, the findings reveal subtle alterations in temporal, frontal, and parietal oscillatory networks during inhibitory processing in ADHD, but effect sizes were modest and embedded within substantial within-group variability. These results support a dimensional view of ADHD neurobiology and highlight the limited discriminative power of network-level EEG markers at the group level.

## Introduction

Attention-deficit/hyperactivity disorder (ADHD) is a neurodevelopmental condition characterized by symptoms of inattention, hyperactivity, and impulsivity that often persist in adulthood and interfere with academic, social, and emotional functioning.[1] Affecting approximately 5 % of children and 2–3 % of adults worldwide, ADHD represents one of the most prevalent and heterogeneous psychiatric disorders. Despite decades of research, the precise neural mechanisms underlying its cognitive and socio-emotional symptoms remain only partly understood.[2–7]

A central feature of ADHD involves deficits in executive functions (EFs), particularly in response inhibition, sustained attention, and performance monitoring. [2,4,8–12] These functions are commonly assessed using Go/NoGo paradigms such as the Visual Continuous Performance Test (VCPT), which reliably differentiates ADHD from neurotypical populations. Typical findings include longer reaction times, increased intra-individual variability, and attenuated event-related potentials (ERPs) such as the N2 and P3 components—markers of conflict detection and inhibitory control.[13–21]

In addition to these executive deficits, individuals with ADHD often exhibit impairments in social cognition, particularly in the recognition and interpretation of emotional facial expressions. Since facial expressions constitute a fundamental channel of non-verbal communication, such deficits can substantially affect interpersonal relationships and emotional regulation.[22–34] Several ERP studies—including our previous work)[35]—have demonstrated reduced amplitudes and delayed latencies of the face-sensitive N170 component in ADHD, suggesting abnormalities in early stages of facial affect processing.[36–41]

However, such traditional ERP approaches, including our earlier study, capture only a limited number of ERP components. Although these components provide valuable information about distinct cognitive subprocesses, they offer merely a fragmented, surface-level view of underlying brain activity. They do not fully reflect the complex spatiotemporal organization of neural processing during the Go/NoGo tasks.

In the present study, we therefore extended this work by applying a network-based analytical approach that enables a more comprehensive characterization of brain activity during cognitive–emotional task performance. Building on the same dataset, we now combine exact low-resolution brain electromagnetic tomography (eLORETA) with functional independent component analysis (fICA). This approach allows us to explore the functional network architecture of the brain during task processing—rather than isolated ERP peaks—by reconstructing the intracortical sources of neural activity beyond the scalp level.

Recent neuroscientific evidence has shown that the brain operates as a highly integrated functional network, in which distributed cortical regions dynamically interact to support executive and emotional processes.[42–49] The eLORETA–fICA framework is ideally suited to capture these large-scale dynamics, as it decomposes

EEG activity into statistically independent networks while localizing their cortical generators. Moreover, this approach enables a detailed examination of frequency-specific neural activity across multiple oscillatory bands.[50–53] Since theta, alpha, beta, and gamma oscillations contribute in distinct ways to cognitive control, emotion regulation, and attentional engagement, analyzing these frequency bands offers deeper insight into how these processes are altered in ADHD compared with healthy controls.[54–66]

Thus, the present study goes significantly beyond our previous ERP analysis by (i) incorporating source-level estimation of neural generators, (ii) examining distributed network interactions rather than isolated components, and (iii) exploring frequency-dependent mechanisms of cognitive–emotional control.

Previous electrophysiological research provides a strong rationale for focusing on theta- and gamma-band dynamics in particular.[55,56,58,60,61,64,67,68] Frontal midline theta oscillations (4–8 Hz) are widely regarded as a core mechanism of cognitive control and conflict monitoring, reflecting top-down regulation and executive engagement.[58,60,66,69,70] Numerous studies have reported reduced frontal theta synchronization and weakened theta coherence in ADHD, consistent with inefficient recruitment of prefrontal control networks during tasks requiring inhibition and sustained attention.[71–73] In contrast, gamma-band oscillations (>30 Hz) have been linked to local cortical excitation, perceptual binding, and emotional salience. Several studies report enhanced gamma activity in ADHD, possibly reflecting compensatory

hyperactivation or cortical hyperarousal associated with reduced large-scale regulatory control.[7,74,75]

Based on this evidence, we formulated the following hypotheses. First, at the conceptual level, we expected frequency- and network-specific alterations in ADHD relative to controls. Specifically, we hypothesized that:

Theta-band activity within fronto-parietal control networks would be reduced in ADHD, reflecting weaker top–down executive engagement during both cognitive (VCPT) and emotional (ECPT) Go/NoGo tasks.

Gamma-band activity within temporal and occipito-temporal regions would be enhanced in ADHD, indicating atypical or compensatory recruitment during facial emotion processing and attentional engagement.

By combining behavioral paradigms with network-based EEG analyses, this study seeks to bridge the gap between ERP-derived component measures and distributed oscillatory dynamics, thereby providing a more comprehensive account of the neural mechanisms linking executive and emotional dysfunction in ADHD.

# Method

*Participants*

The dataset analyzed in this study corresponds to the same multicenter ADHD project described in earlier publications from our group. From this cohort, a total of 272 participants fulfilled all inclusion criteria and completed both the VCPT and ECPT tasks (ADHD: n = 102; controls: n = 170). Participants ranged in age from 6 to 60 years and were assigned to the categories children, adolescents, and adults (Table 1). ADHD diagnoses had been established by trained psychiatrists and clinical psychologists based on DSM-5 criteria. Exclusion criteria comprised neurological disorders, traumatic brain injury, epilepsy, substance misuse, pregnancy, insufficient German language proficiency, and an IQ below 80. All participants were unmedicated on the testing day.

**Please place Table 1 here.**

Go/NoGo Tasks

Two structurally parallel Go/NoGo paradigms were administered: the Visual Continuous Performance Test (VCPT) and the Emotional Continuous Performance Test (ECPT). The VCPT comprised images of animals, plants, and humans. Participants were instructed to respond whenever an animal stimulus was preceded by another animal (Go), to withhold responses when an animal was followed by a plant (NoGo), and to ignore all other combinations, which served as neutral trials and were accompanied by a brief auditory tone.

The ECPT mirrored the structure of the VCPT but used photographs of male and female actors depicting neutral, happy, and angry facial expressions. In this task, an angry–angry sequence constituted a Go trial, while angry–happy sequences served as NoGo trials. Neutral and happy–happy sequences were defined as Ignore trials and were also paired with an auditory tone. Each task comprised 100 trials per condition. Stimuli were presented for 100 ms, with the first stimulus appearing at 300 ms and the second at 1400 ms after trial onset, followed by a fixed intertrial interval of 3000 ms. The total duration for each task was approximately 22 minutes.

**Please place Figure 1 here.**

*EEG Recording and Preprocessing*

EEG data were recorded using a 23-channel NeuroAmp® x23 DC-coupled system (BEE Medic GmbH, Switzerland) with a nominal sampling rate of 500 Hz, which was subsequently downsampled to 250 Hz. Signals were band-pass filtered between .5 and 50 Hz and re-referenced to the common average. Nineteen tin electrodes embedded in an EEG cap (Electro-Cap International Inc., USA) were positioned according to the international 10–20 system, and impedance was kept below 5 kΩ.

Preprocessing followed the procedures applied in the baseline investigation. Eye blinks and horizontal eye movements were identified using independent component analysis (ICA), and components associated with ocular artifacts were removed. Segments with excessively low-frequency (0–3 Hz) or high-frequency (20–50 Hz) activity, as well as epochs exceeding ±100 µV, were rejected. The resulting artifact-free data were processed using custom-built EEGLAB and ERPrec functions implemented in MATLAB.

*EEG Source Reconstruction and Functional Independent Component Analysis*

Source localization was carried out using exact low-resolution brain electromagnetic tomography (eLORETA), which estimates cortical current density within a solution space of 6239 gray-matter voxels (5-mm resolution; MNI152 template). Spectral density images were computed for eight canonical frequency bands: delta (1.5–6 Hz), theta (6.5–8 Hz), alpha-1 (8.5–10 Hz), alpha-2 (1.5–12 Hz), beta-1 (12.5–18 Hz), beta-2 (18.5–21 Hz), high-beta (22–30 Hz), and gamma (30–50 Hz). Each participant, therefore, contributed eight frequency-specific cortical maps for each task and condition.

Following the procedures described by Frei et al. and Aoki et al.,[51,53,76] these frequency-band images were concatenated into participant-level feature matrices and submitted to functional independent component analysis (fICA). Statistical independence was optimized using fourth-order cumulants. For all analyses, ten independent components (ICs) were retained, representing the dominant spectral-spatial functional networks across tasks and conditions. ICs were sorted by explained variance, and spatial maps were thresholded at $z = 3$. Warm colors indicated relative increases, and cool colors relative decreases, in spectral power as a function of IC loading strength.

Each participant received ten IC loading coefficients for each of the four conditions (VCPT-Go, VCPT-NoGo, ECPT-Go, ECPT-NoGo), which served as dependent variables in the statistical analyses.

**Statistical Analysis**

Statistical analyses were performed on the IC loading coefficients derived from the fICA decomposition. Mixed-model ANCOVAs were computed with Group (ADHD vs. control) as a between-subject factor and Task (VCPT vs. ECPT) and Condition (Go vs. NoGo) as within-subject factors. Age and sex were entered as covariates, and participant ID served as a random factor. Analyses were carried out using the afex package in R, and generalized eta squared ($η^2\_G$) was used as the effect-size metric.

Behavioral performance measures (including mean reaction time, reaction-time variability, commission errors, and omission errors) were extracted for both Go/NoGo tasks and matched to the EEG sample by participant ID. These measures were analyzed using the same ANCOVA model structure (Group × Task, with Age and Sex as covariates). Post-hoc comparisons were computed with estimated marginal means and Holm correction.

Although gamma-band activity (>30 Hz) is potentially influenced by high-frequency myogenic noise, it was retained to provide a complete spectral overview; interpretations of gamma-related findings were made with appropriate caution.

**Results**

The eLORETA–ICA decomposition applied to the Go/NoGo EEG data yielded ten independent components (ICs) that were stable across both tasks (Figure 2). Each IC was characterized by a distinct spatial pattern and dominant oscillatory profile, spanning posterior occipito-temporal visual cortices, inferior frontal cortex, parietal association areas, medial parietal regions, lateral and superior temporal cortices, and fronto-central midline regions. Dominant frequencies ranged from delta to gamma, with the majority of components showing peak activation in higher-frequency bands.

Across tasks and conditions, several components showed reliable group differences (Table 2). ADHD participants exhibited overall stronger activation than controls in four networks: IC1 (medial frontal gamma), IC3 (inferior temporal gamma), IC4 (occipital alpha-2), and IC7 (medial frontal alpha-1). These effects were consistent across both the VCPT and ECPT and reflect broader hyperactivation tendencies in ADHD across widespread large-scale cortical networks.

Among the mentioned components showing group differences, IC3, localized to the bilateral inferior temporal gyrus with a dominant gamma-band profile, showed the clearest group-related differences (Figure 2, 3). A significant main effect of Group (ADHD > controls; $F(1,268) = 9.49$, $p \leq .01$, $\eta^2\_G = .02$) was accompanied by a significant Group × Condition interaction ($F(1,268) = 8.82$, $p \leq .01$). Post-hoc contrasts revealed disproportionately stronger IC3 activation in ADHD during NoGo trials, whereas Go-related activity was largely comparable across groups ($d = .3$). This pattern indicates a subtle but reliable enhancement of temporal gamma recruitment during inhibitory demands in ADHD.

**Please place Figures 2 and 3 here**

Additionally, significant Group × Condition interactions were observed in IC6 and IC9 with marginal effect sizes (Table 2). These interactions reflected disproportionately increased NoGo-related activation in ADHD, suggesting heightened or inefficient recruitment of bilateral inferior temporal gamma (IC3), fronto-central delta (IC6), and temporo-parietal gamma (IC9) networks during inhibitory events. The fronto-central delta interaction in IC6 showed a clear pattern of enhanced NoGo activation in ADHD, consistent across both tasks. Post-hoc contrasts confirmed that ADHD showed higher activation than controls in all significant interactions. For IC3, ADHD >

control differences were larger during NoGo (Control–ADHD = –4022, p < .001) than Go (–2887, p = .01). A similar NoGo-specific increase was observed for IC6 (–1320, p = .01) and IC9 which missed the significance threshold after the posthocs (–1123, p = .21).

No Group × Task or Group × Task × Condition interactions were observed, indicating that group-related neural differences in the processing of Go versus NoGo events remained stable across emotional and non-emotional task contexts.

A main effect of Task was observed only for IC8 (middle temporal delta), which showed higher activation during the VCPT compared to the ECPT (VCPT–ECPT = 857, p = .007, d = 0.33). This indicates a task-dependent modulation of temporal delta activity independent of diagnostic group. No other components demonstrated task-selective modulation, indicating that emotional content selectively influenced temporal delta activity but did not substantially alter activation in most large-scale cortical networks.

Two additional effects reached significance and exhibited very small effect sizes. IC10 showed a weak Task × Condition interaction ($\eta^2\_G$ = .0003), with slightly higher activation during the VCPT than the ECPT in both Go and NoGo trials (VCPT–ECPT = +1024 and +917, respectively; both p < .02). Given the minimal effect size and absence of a theoretical prediction, this result is reported for completeness only.

**Please place Table 2 here**

Behavioural analyses (Figure 4) closely paralleled the neural findings and reproduced the established ADHD performance profile (Tables 3 and 4). Across both tasks, ADHD participants responded more slowly, exhibited substantially greater reaction-time variability, and committed more commission and omission errors than controls. Strong task effects were observed for all behavioural outcomes, with the ECPT producing slower and more variable responses as well as increased error rates in both groups. Significant Group × Task interactions for reaction time, reaction-time variability, and omission errors indicated that the emotional context of the ECPT disproportionately amplified ADHD-related deficits. Descriptive means and variances for all behavioural measures are reported in Table 3.

**Please place Tables 3 and 4 here**

**Please place Figure 4 here**

## Discussion

The present study investigated large-scale cortical network dynamics underlying cognitive and facial-emotional Go/NoGo performance in ADHD using an eLORETA–ICA framework combined with behavioural metrics. Across both tasks, we extracted ten stable independent components spanning posterior visual, temporal, parietal, and fronto-central regions. Several components showed statistically reliable between-group differences, and one component (IC3) demonstrated a consistent Group × Condition interaction reflecting altered NoGo-related recruitment in ADHD. As we expected, in behavioural measures, the ADHD participants exhibited the well-documented pattern of slower and more variable responses as well as higher rates of commission and omission errors, particularly during the emotional ECPT.[11,19,77–80] Together, these findings provide an integrated behavioural–neural account of cognitive and emotional control in ADHD, extending previous ERP results from the same dataset.[35]

Consistent with large-scale EEG research on ADHD, the present results reveal statistically significant, but minimal, group differences in distributed cortical networks.[20,35,81–83] One illustrative example is IC3 (figure 3), a gamma-dominant network centred in the inferior temporal gyrus, which showed both a small Group effect and a Group × Condition interaction. The inferior temporal gyrus is involved in higher-order visual processing, perceptual discrimination, and facial-emotional interpretation,[84–86] and gamma oscillations in this region are typically associated with local cortical excitation and salience-related processing as well as a mechanism for managing multi-item working memory.[60,64,87–90] The stronger NoGo-related IC3 activation in ADHD may therefore reflect subtle compensatory engagement of perceptual–attentional mechanisms when inhibitory control is required. However, the effect size was modest ($\eta^2\_G \approx .02$), and substantial within-group variability remained, underscoring that such network-level alterations represent subtle modulations rather than discrete neural signatures of ADHD.

Across the ten ICA-derived components, diagnostic status explained only a few percent of variance. In most cases, age—and to a lesser extent, sex—exerted a more substantial influence on network activity than Group. This closely parallels the findings of Münger et al. (2021),[20] who analysed a large (n = 674) ERP dataset from the same multicentre ADHD project and likewise observed statistically significant but small-to-moderate effects, with substantial (≈80%) overlap between ADHD and control distributions even for moderate effect sizes. Similar patterns were also seen in our previous ERP study using an overlapping subset of the cohort, where robust behavioural group differences co-occurred with only modest neural effects.[35] Together, these converging results across ERP and source-level analyses indicate that ADHD-related neural alterations are detectable but substantially smaller than behavioural differences and far below what would be expected under a categorical disease model.

Rather than reflecting the absence of meaningful neurocognitive variability, these findings reinforce that ADHD does not manifest as a strongly discriminating neural subtype. The modest magnitude of neural effects is consistent with the substantial heterogeneity and diagnostic imprecision inherent to clinically defined ADHD—a point frequently discussed in the literature.[2,6,81–83,91–100] ADHD groups comprise individuals with diverse cognitive profiles, symptom constellations, developmental trajectories, and environmental or therapeutic histories. Such heterogeneity blurs group boundaries and reduces separability across neural measures. Long-term medication exposure, behavioural therapy, school interventions, and developmental compensation may further contribute to overlapping patterns of cortical activation.[20,81–83,98]

Taken together, the present results suggest that large-scale cortical network organization during cognitive and emotional Go/NoGo processing is shared mainly between individuals with and without ADHD, with only subtle modulations in specific networks, including posterior alpha–gamma components and temporal delta generators. These effects were statistically reliable but small, highlighting the importance of interpreting ADHD-related EEG differences within a dimensional, variance-sensitive framework rather than a categorical neuropathological model. This perspective aligns with contemporary theories suggesting that cognitive and affective dysfunctions in ADHD arise from distributed and relatively subtle network-level shifts rather than from discrete pathological circuits.[10,20,94–97,101–104]

## Limitations

Several limitations should be considered when interpreting these results. First, although eLORETA–ICA provides a robust data-driven framework for decomposing large-scale networks, the spatial precision of EEG source localization is inherently limited, and the exact anatomical boundaries of each IC should be interpreted cautiously, particularly given the minor group effects. Second, the ADHD sample reflects the heterogeneity of clinical practice: symptom severity, developmental histories, comorbidities, and prior medication exposure were not controlled experimentally and are likely to have contributed to the significant within-group variability. Third, while the sample size is relatively large for an EEG network study, it remains modest relative to the heterogeneity of ADHD, which limits the identification of potential subgroups. Fourth, the wide age range, though statistically controlled, introduces strong developmental effects that may blur or interact with diagnostic differences.

Finally, behavioural performance and neural activity were assessed within a laboratory Go/NoGo paradigm, which captures only specific aspects of inhibitory and emotional processing. Real-world executive functioning, emotion recognition, and attentional control are more complex and may rely on additional network dynamics not optimally captured by the current design.

## Conclusion

This study provides source-level evidence that ADHD-related alterations in cognitive and emotional Go/NoGo processing are consistent but small, embedded within otherwise largely shared large-scale cortical network architectures. Although certain components showed significant group and interaction effects, these accounted for only a small proportion of the variance and were overshadowed by developmental and demographic influences. Behavioural impairments were large and robust, but they were not accompanied by comparably strong neural differentiations. Together, these findings support a dimensional, variance-sensitive view of ADHD neurobiology and highlight the limitations of relying on categorical diagnostic labels to identify discrete neural biomarkers. Future studies should prioritize stratification approaches, developmental modelling, and individual-differences frameworks to capture better the heterogeneity underlying ADHD.

## Author contributions

SV processed the experimental data, performed the analysis, and drafted the manuscript. GC collected data, processed the experimental data, and reviewed the manuscript. JK and HAR collected data. DE and AM designed the project. LJ supervised the project and contributed to writing the manuscript and interpreting the results.

## Funding

The authors disclose the following financial support for the research, authorship, and/or publication of this article: This work was supported by the Uniscientia Foundation and the Brain and Trauma Foundation Grison Switzerland.

## Acknowledgments

We would like to extend our sincere thanks to the participants in this study and the clinical staff involved in the data collection.

We acknowledge the use of OpenAI for assistance with language refinement, including rephrasing and synonym suggestions, in the preparation of this manuscript.

## Declaration of Conflicting Interests

The authors declared no potential conflicts of interest regarding the research, authorship, and/or publication of this article.

## Ethical Approval

# Tables

*Table 1. Descriptive information of the participating subjects (n=272). The number of male and female subjects is shown. In addition, the mean age and IQ (and standard deviation) for both groups are presented.*

|  | ADHD | Control |
|---|---|---|
| N total | 102 | 170 |
| Male (N%) | 65 (63.7%) | 67 (39.4%) |
| Female (N%) | 37 (36.3%) | 103 (6.6%) |
| Age (mean ± SD) | 28.55 ± 19.24 | 24.24 ± 14.28 |
| IQ | 102.81 ± 11.98 | 107.53 ± 11.15 |

*Table 2. Repeated-measures ANCOVAs for each independent component (IC1–IC10) with Group (ADHD vs Control) as a between-subjects factor and Task (VCPT, ECPT) and Condition (Go, NoGo) as within-subjects factors. Age and Sex were covariates. Table shows Df, F, p and generalized eta-squared (ges).*

| IC | Effect | Df | F_value | p_value | ges |
|---|---|---|---|---|---|
| IC1 | Group | 1 | 7.36 | **≤ .01** | .01 |
|  | Sex | 1 | 2.29 | .13 | ≤ .01 |
|  | Age | 1 | .22 | .63 | ≤ .001 |
|  | Task | 1 | 1.27 | .25 | ≤ .001 |
|  | Group × Task | 1 | .15 | .69 | ≤ .001 |
|  | Sex × Task | 1 | .001 | .96 | ≤ .001 |
|  | Age × Task | 1 | .02 | .86 | ≤ .001 |
|  | Condition | 1 | 2.16 | .14 | ≤ .001 |
|  | Group × Condition | 1 | 3.12 | .07 | ≤ .001 |
|  | Sex × Condition | 1 | .36 | .54 | ≤ .001 |
|  | Age × Condition | 1 | 1.60 | .20 | ≤ .001 |
|  | Task × Condition | 1 | .0004 | .98 | ≤ .001 |
|  | Group × Task × Condition | 1 | .003 | .95 | ≤ .001 |
| IC2 | Group | 1 | .84 | .35 | ≤ .01 |
|  | Sex | 1 | .008 | .92 | ≤ .001 |
|  | Age | 1 | 1.21 | .27 | ≤ .01 |
|  | Task | 1 | .33 | .56 | ≤ .001 |
|  | Group × Task | 1 | .12 | .72 | ≤ .001 |
|  | Sex × Task | 1 | 5.08 | **≤ .05** | ≤ .01 |
|  | Age × Task | 1 | .38 | .53 | ≤ .001 |
|  | Condition | 1 | 1.07 | .29 | ≤ .01 |
|  | Group × Condition | 1 | .05 | .80 | ≤ .001 |
|  | Sex × Condition | 1 | .29 | .58 | ≤ .001 |
|  | Age × Condition | 1 | .27 | .59 | ≤ .001 |
|  | Task × Condition | 1 | 2.40 | .12 | ≤ .01 |
|  | Group × Task × Condition | 1 | .21 | .64 | ≤ .001 |
| IC3 | Group | 1 | 9.49 | **≤ .01** | .02 |

|  | Effect | df | F | p | η² |
|---|---|---|---|---|---|
|  | Sex | 1 | .01 | .91 | ≤ .001 |
|  | Age | 1 | 2.65 | **≤ .001** | .06 |
|  | Task | 1 | 1.60 | .20 | ≤ .001 |
|  | Group × Task | 1 | .16 | .68 | ≤ .001 |
|  | Sex × Task | 1 | .01 | .91 | ≤ .001 |
|  | Age × Task | 1 | .23 | .62 | ≤ .001 |
|  | Condition | 1 | .80 | .36 | ≤ .001 |
|  | Group × Condition | 1 | 8.82 | **≤ .01** | ≤ .001 |
|  | Sex × Condition | 1 | .03 | .85 | ≤ .001 |
|  | Age × Condition | 1 | 2.11 | .14 | ≤ .001 |
|  | Task × Condition | 1 | .04 | .82 | ≤ .001 |
|  | Group × Task × Condition | 1 | .42 | .51 | ≤ .001 |
| IC4 | Group | 1 | 4.10 | **≤ .05** | .01 |
|  | Sex | 1 | .72 | .39 | ≤ .01 |
|  | Age | 1 | .43 | .50 | ≤ .01 |
|  | Task | 1 | .16 | .68 | ≤ .001 |
|  | Group × Task | 1 | 3.03 | .08 | ≤ .01 |
|  | Sex × Task | 1 | .51 | .47 | ≤ .001 |
|  | Age × Task | 1 | 1.32 | .25 | ≤ .001 |
|  | Condition | 1 | .02 | .87 | ≤ .001 |
|  | Group × Condition | 1 | .41 | .52 | ≤ .001 |
|  | Sex × Condition | 1 | .02 | .86 | ≤ .001 |
|  | Age × Condition | 1 | 1.33 | .24 | ≤ .001 |
|  | Task × Condition | 1 | ≤ .001 | .99 | ≤ .001 |
|  | Group × Task × Condition | 1 | .24 | .61 | ≤ .001 |
| IC5 | Group | 1 | 1.73 | .18 | ≤ .001 |
|  | Sex | 1 | 1.57 | .21 | ≤ .001 |
|  | Age | 1 | 5.50 | **≤ .01** | .01 |
|  | Task | 1 | .62 | .42 | ≤ .001 |
|  | Group × Task | 1 | .15 | .69 | ≤ .001 |
|  | Sex × Task | 1 | .004 | .94 | ≤ .001 |
|  | Age × Task | 1 | .68 | .40 | ≤ .001 |
|  | Condition | 1 | 3.84 | **≤ .05** | ≤ .001 |
|  | Group × Condition | 1 | 3.06E-06 | .99 | ≤ .001 |
|  | Sex × Condition | 1 | .34 | .55 | ≤ .001 |
|  | Age × Condition | 1 | .01 | .89 | ≤ .001 |
|  | Task × Condition | 1 | .07 | .78 | ≤ .001 |
|  | Group × Task × Condition | 1 | .01 | .88 | ≤ .001 |
| IC6 | Group | 1 | 1.82 | .17 | ≤ .001 |
|  | Sex | 1 | 2.17 | .14 | ≤ .001 |
|  | Age | 1 | 2.72 | .09 | ≤ .001 |
|  | Task | 1 | .71 | .39 | ≤ .001 |
|  | Group × Task | 1 | .15 | .69 | ≤ .001 |
|  | Sex × Task | 1 | 3.15 | .07 | ≤ .001 |
|  | Age × Task | 1 | .01 | .88 | ≤ .001 |
|  | Condition | 1 | 1.22 | .26 | ≤ .001 |
|  | Group × Condition | 1 | 16.18 | **≤ .001** | ≤ .01 |
| | Sex × Condition | 1 | 1.48 | .22 | ≤ .001 |
| | Age × Condition | 1 | 9.10 | **≤ .01** | ≤ .01 |
| | Task × Condition | 1 | 1.60 | .20 | ≤ .001 |
| | Group × Task × Condition | 1 | .51 | .47 | ≤ .001 |
| IC7 | Group | 1 | 4.87 | .02 | .01 |
| | Sex | 1 | 7.65 | **≤ .01** | .02 |
| | Age | 1 | 3.92 | **≤ .05** | .01 |
| | Task | 1 | .87 | .35 | ≤ .001 |
| | Group × Task | 1 | .43 | .51 | ≤ .001 |
| | Sex × Task | 1 | 1.51 | .21 | ≤ .001 |
| | Age × Task | 1 | .25 | .61 | ≤ .001 |
| | Condition | 1 | 4.49 | **≤ .05** | ≤ .001 |
| | Group × Condition | 1 | .21 | .64 | ≤ .001 |
| | Sex × Condition | 1 | 1.81 | .17 | ≤ .001 |
| | Age × Condition | 1 | 1.27 | .26 | ≤ .001 |
| | Task × Condition | 1 | 2.67 | .10 | ≤ .001 |
| | Group × Task × Condition | 1 | .07 | .79 | ≤ .001 |
| IC8 | Group | 1 | .77 | .37 | ≤ .01 |
| | Sex | 1 | 1.07 | .29 | ≤ .01 |
| | Age | 1 | 4.10 | **≤ .05** | .01 |
| | Task | 1 | 5.05 | **≤ .05** | ≤ .001 |
| | Group × Task | 1 | .06 | .79 | ≤ .001 |
| | Sex × Task | 1 | 1.20 | .27 | ≤ .001 |
| | Age × Task | 1 | .83 | .36 | ≤ .001 |
| | Condition | 1 | 2.67 | .10 | ≤ .001 |
| | Group × Condition | 1 | .43 | .51 | ≤ .001 |
| | Sex × Condition | 1 | .0003 | .98 | ≤ .001 |
| | Age × Condition | 1 | .08 | .77 | ≤ .001 |
| | Task × Condition | 1 | .48 | .48 | ≤ .001 |
| | Group × Task × Condition | 1 | 1.05 | .30 | ≤ .001 |
| IC9 | Group | 1 | .76 | .38 | ≤ .001 |
| | Sex | 1 | 1.18 | .27 | ≤ .001 |
| | Age | 1 | 1.68 | .19 | ≤ .001 |
| | Task | 1 | 1.10 | .29 | ≤ .001 |
| | Group × Task | 1 | 1.44 | .23 | ≤ .001 |
| | Sex × Task | 1 | .12 | .72 | ≤ .001 |
| | Age × Task | 1 | 5.16 | **≤ .05** | ≤ .001 |
| | Condition | 1 | .38 | .53 | ≤ .001 |
| | Group × Condition | 1 | 4.95 | **≤ .05** | ≤ .001 |
| | Sex × Condition | 1 | .51 | .47 | ≤ .001 |
| | Age × Condition | 1 | 6.11 | **≤ .01** | ≤ .001 |
| | Task × Condition | 1 | .31 | .57 | ≤ .001 |
| | Group × Task × Condition | 1 | .83 | .36 | ≤ .001 |
| IC10 | Group | 1 | 1.88 | .17 | ≤ .01 |
| | Sex | 1 | 1.69 | .19 | ≤ .01 |
| | Age | 1 | 4.65 | **≤ .05** | .01 |
| | Task | 1 | 1.17 | .27 | ≤ .001 |
| | Group × Task | 1 | .76 | .38 | ≤ .001 |
| | Sex × Task | 1 | .0001 | .99 | ≤ .001 |
| | Age × Task | 1 | .43 | .51 | ≤ .001 |
| | Condition | 1 | 2.25 | .13 | ≤ .001 |
| | Group × Condition | 1 | 1.73 | .18 | ≤ .001 |
| | Sex × Condition | 1 | .01 | .90 | ≤ .001 |
| | Age × Condition | 1 | .80 | .37 | ≤ .001 |
| | Task × Condition | 1 | 4.16 | **≤ .05** | ≤ .001 |
| | Group × Task × Condition | 1 | .47 | .49 | ≤ .001 |

*Table 3. Descriptive behavioural performance (mean ± SD) for VCPT and ECPT in ADHD and control groups. RT = reaction time; RTvar = reaction-time variability.*

| | Control VCPT | ADHD VCPT | Control ECPT | ADHD ECPT |
|---|---|---|---|---|
| Commission | .55 ± 1.01 | 1.17 ± 2.26 | 1.83 ± 2.53 | 3.10 ± 3.68 |
| Omission | 3.02 ± 4.48 | 8.32 ± 9.79 | 9.11 ± 11.30 | 21.98 ± 19.04 |
| RT mean | 368.25 ± 9.56 | 405.42 ± 9.88 | 415.27 ± 94.57 | 467.28 ± 98.71 |
| RTvar | 8678.90 ± 7307.08 | 15327.46 ± 9224.72 | 13517.51 ± 92.26 | 22154.03 ± 10751.30 |

*Table 4. ANCOVA results for behavioural measures (RTmean = reaction time, RTvar = reaction-time variability, Commission and Omission errors) across VCPT and ECPT tasks, testing Group, Task, and Group × Task effects while controlling for Sex and Age. ges = generalized eta squared.*

| | Effect | Df | F | p | ges |
|---|---|---|---|---|---|
| RT mean | Group | 1 | 16.20 | **≤ .001** | .05 |
| | Sex | 1 | .71 | .39 | ≤ .01 |
| | Age | 1 | 11.88 | **≤ .001** | .04 |
| | Group × Sex | 1 | .33 | .56 | ≤ .001 |
| | Task | 1 | 399.37 | **≤ .001** | .07 |
| | Group × Task | 1 | 9.35 | **≤ .01** | .001 |
| | Sex × Task | 1 | .47 | .49 | ≤ .001 |
| | Age × Task | 1 | 2.77 | .09 | ≤ .001 |
| RTvar | Group | 1 | 59.72 | **≤ .001** | .15 |
| | Sex | 1 | 1.22 | .26 | ≤ .001 |
| | Age | 1 | 2.13 | **≤ .001** | .05 |
| | Group × Sex | 1 | .001 | .96 | ≤ .001 |
|  | Task | 1 | 14.35 | **≤ .001** | .09 |
|  | Group × Task | 1 | 5.78 | **≤ .01** | ≤ .001 |
|  | Sex × Task | 1 | 1.77 | .18 | ≤ .001 |
|  | Age × Task | 1 | .61 | .43 | ≤ .001 |
| Commission | Group | 1 | 12.04 | **≤ .001** | .03 |
|  | Sex | 1 | .93 | .33 | ≤ .001 |
|  | Age | 1 | 1.52 | .21 | ≤ .001 |
|  | Group × Sex | 1 | 1.27 | .26 | ≤ .001 |
|  | Task | 1 | 98.0 | **≤ .001** | .08 |
|  | Group × Task | 1 | 2.09 | .14 | ≤ .001 |
|  | Sex × Task | 1 | 7.89 | **≤ .01** | ≤ .001 |
|  | Age × Task | 1 | .75 | .38 | ≤ .001 |
| Omission | Group | 1 | 49.62 | **≤ .001** | .12 |
|  | Sex | 1 | 2.29 | .13 | ≤ .001 |
|  | Age | 1 | 9.16 | **≤ .01** | .02 |
|  | Group × Sex | 1 | .0003 | .98 | ≤ .001 |
|  | Task | 1 | 197.27 | **≤ .001** | .14 |
|  | Group × Task | 1 | 26.84 | **≤ .001** | .02 |
|  | Sex × Task | 1 | 4.44 | **≤ .05** | ≤ .001 |
|  | Age × Task | 1 | 8.55 | **≤ .01** | ≤ .001 |

# Figures

*Figure 1.* VCPT and ECPT conditions, including Go, NoGo, and distractor/ignore.

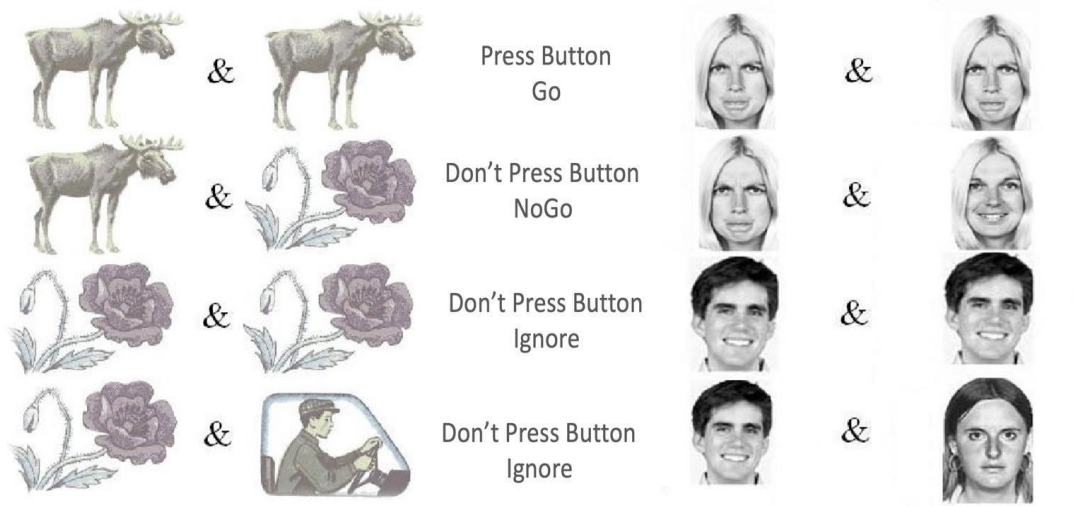

*Figure 2.* Shown are the ten independent components (IC1–IC10) extracted from the full sample across both tasks. Red/yellow areas indicate regions where IC activity increases with higher component coefficients, whereas blue areas represent relative decreases. Each IC reflects a stable large-scale cortical network used for all subsequent statistical analyses.

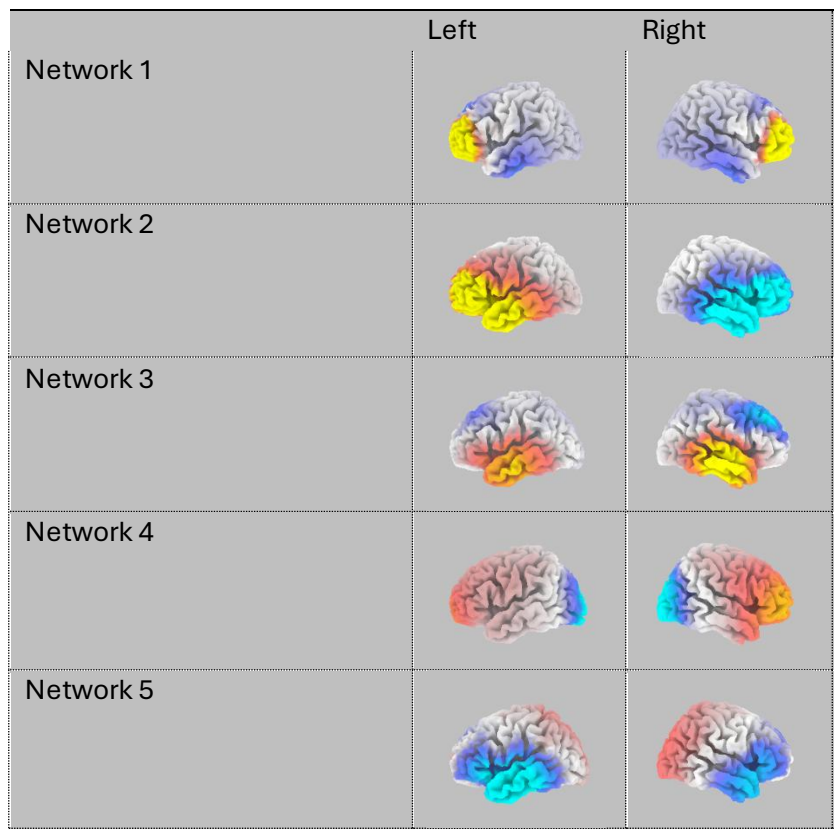

| | | |
|---|---|---|
| Network 6 | 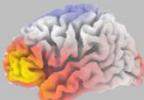 | 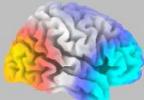 |
| Network 7 | 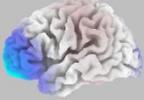 | 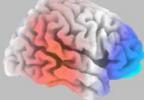 |
| Network 8 | 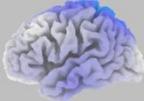 | 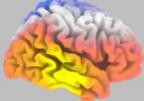 |
| Network 9 | 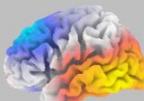 | 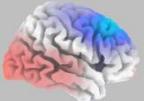 |
| Network 10 | 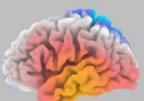 | 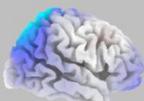 |

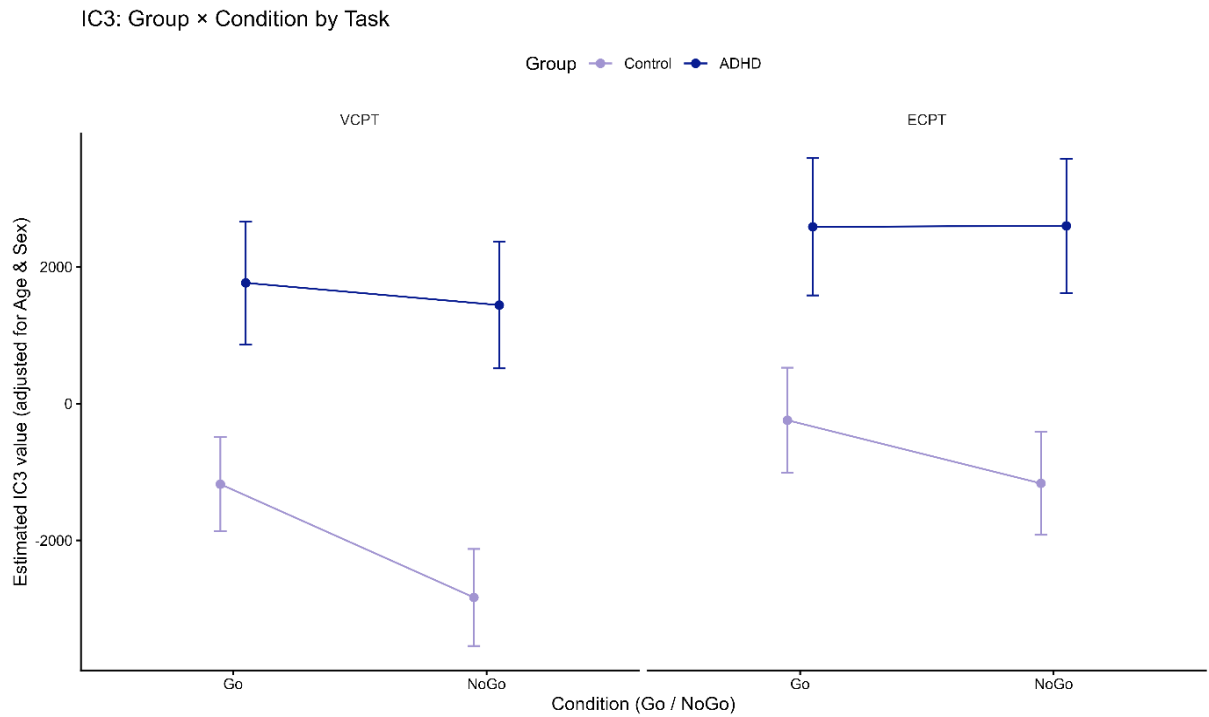

*Figure 3.* IC3 (inferior temporal gamma network): Estimated marginal means depicting a significant Group × Condition interaction. ADHD participants show disproportionately increased IC3 activation during NoGo trials compared with controls, consistent across VCPT and ECPT.

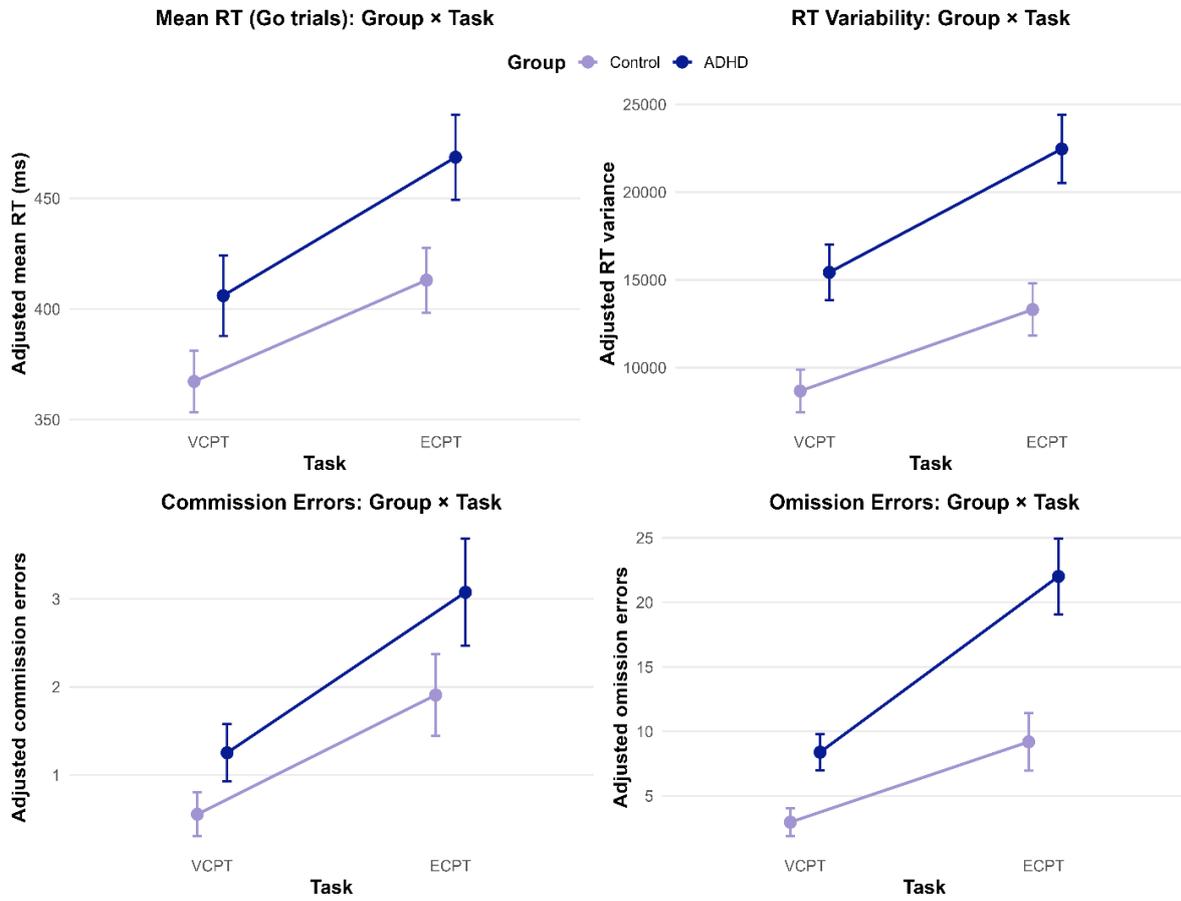

*Figure 4.* Behavioural group differences across tasks showing *adjusted means (±95% CI) for* **RT, RT variability, commission errors,** *and* **omission errors** *in* **VCPT** *and* **ECPT**. *ADHD participants show slower and more variable responses and higher error rates, with larger deficits in the ECPT.*